\pdfoutput=1
\documentclass[sigplan,10pt]{acmart}

\makeatletter                   %
\def\mdseries@tt{m}             %
\makeatother                    %

\acmConference[To Appear in PACT'18]{}{November 1--4, 2018}{Limassol, Cyprus}

\acmYear{2018}
\acmISBN{} %
\acmDOI{} %
\startPage{1}

\setcopyright{none}

\usepackage{booktabs}   %
\usepackage{subcaption} %
\usepackage[amssymb]{SIunits}            %
\usepackage{sistyle}
\usepackage{courier}            %
\usepackage{alltt} %
\usepackage{url}                  %
\usepackage{listings}          %

\usepackage[frozencache=true]{minted}          %

\usepackage{enumitem}      %

\usepackage{color} %
\usepackage{graphics}
\usepackage{graphicx}
\usepackage{multirow}
\usepackage{caption}

\usepackage{breakurl} %
\usepackage{references}

\usepackage{multirow}
\usepackage{array}
\usepackage{paralist}

\usepackage{gitinfo}

\begin{document}

\title[Cimple]{Cimple: Instruction and Memory Level Parallelism
}         %
\subtitle{A DSL for Uncovering ILP and MLP }                     %

\author{Vladimir Kiriansky,\  Haoran Xu,\  Martin Rinard,\  Saman Amarasinghe}
\affiliation{
  \institution{MIT CSAIL}            %
}
\email{{vlk, haoranxu510, rinard, saman}@csail.mit.edu}

\newcommand{\added}[1]{%
  \cbcolor{red}
  \begin{changebar}
    #1
  \end{changebar}%
  }%

\newcommand\TODO[1]{\mbox{\textcolor{red}{#1}}}

\newcommand\Cimplensp{Cimple{}}
\newcommand\Cimplesp{Cimple{ }}
\newcommand\CIMPLE{\textsc{Cimple}{}}
\newcommand\CIMPLEnsp{\CIMPLE}
\newcommand\CIMPLEsp{\CIMPLE{ }}

\newcommand{\Syntax}[1]{{\textbf{\texttt{#1}}}}
\newcommand{\SyntaxFirst}[1]{{\textbf{\texttt{#1}}}}
\newcommand{\SyntaxTail}[1]{{\texttt{#1}}}
\newcommand{\keyword}[1]{{\texttt{#1}}}
\newcommand{\Stmt}[1]{{\textbf{\texttt{#1}}}}
\newcommand{\mytimes}{$\times$}

\newcommand\IMLP{{IMLP} {}}
\newcommand\IMLPsp{{IMLP} {}}
\newcommand\IMLPnsp{{IMLP}}

\newcommand\naive{na\"ive{ }}

\newcommand{\floor}[1]          {\left\lfloor #1 \right\rfloor}
\newcommand{\ceil}[1]           {\left\lceil #1 \right\rceil}
\newcommand{\tuple}[1]            {\ifmmode{\left\langle #1 \right\rangle}
                                 \else{$\left\langle${#1}$\right\rangle$}\fi}
\newcommand{\revision}{rev \gitAbbrevHash}

\definecolor{linenumcolor}{rgb}{0.5,0.5,1.0}
\definecolor{myschedule}{rgb}{0.858, 0.188, 0.478}

\renewcommand{\theFancyVerbLine}{\sffamily
  \textcolor{linenumcolor}{\scriptsize
    \oldstylenums{\arabic{FancyVerbLine}}}}

\begin{abstract}

Modern out-of-order processors have increased capacity to exploit
instruction level parallelism (ILP) and memory level parallelism
(MLP), e.g., by using wide superscalar pipelines and vector execution
units, as well as deep buffers for in-flight memory requests.  These
resources, however, often exhibit poor utilization rates on workloads
with large working sets, e.g., in-memory databases, key-value stores,
and graph analytics, as compilers and hardware struggle to expose ILP
and MLP from the instruction stream automatically.

In this paper, we introduce the \textbf{IMLP} (Instruction and Memory Level
Parallelism) task programming model.  IMLP tasks execute as coroutines that
yield execution at annotated long-latency operations, e.g., memory
accesses, divisions, or unpredictable branches.  IMLP tasks are
interleaved on a single thread, and integrate well with thread
parallelism and vectorization.
Our DSL embedded in C++, Cimple, allows
exploration of task scheduling and transformations, such as
buffering, vectorization, pipelining, and prefetching.

  We demonstrate state-of-the-art performance on core
algorithms used in in-memory databases that operate on arrays, hash
tables, trees, and skip lists.  Cimple applications reach 
2.5$\times$ throughput gains over hardware multithreading on a multi-core, and 6.4$\times$  
single thread speedup.

\end{abstract}

\maketitle
\renewcommand{\shortauthors}{Kiriansky et al.}

\section{Introduction}
\label{sec:intro}

Barroso et al.~\cite{barroso17cacm-killer-us} observe that ``killer
microseconds'' prevent efficient use of modern datacenters.  The
critical gap between millisecond and nanosecond latencies is outside
of the traditional roles of software and hardware.  Existing software
techniques used to hide millisecond latencies, such as threads or
asynchronous I/O, have too much overhead when addressing microsecond
latencies and below.  On the other hand, out-of-order hardware is capable
of hiding at most tens of nanoseconds latencies.  Yet, average memory access times now
span a much broader range: from \verb|~|20~ns for L3 cache hits, to
$>$200~ns for DRAM accesses on a remote NUMA node --- making hardware
techniques inadequate.  We believe an efficient, flexible, and
expressive programming model can scale the full memory hierarchy from tens to hundreds of nanoseconds.

Processors have grown their capacity to exploit instruction level
parallelism (ILP) with wide scalar and vector pipelines, e.g., cores
have 4-way superscalar pipelines, and vector units can execute 32
arithmetic operations per cycle.  Memory level parallelism (MLP) is
also pervasive with deep buffering between caches and DRAM that allows
10+ in-flight memory requests per core.  Yet, modern CPUs
still struggle to extract matching ILP and MLP from the program
stream.

Critical infrastructure applications, e.g., in-memory databases,
key-value stores, and graph analytics, characterized by large working
sets with multi-level address indirection and pointer traversals push
hardware to its limits: large multi-level caches and
branch predictors fail to keep processor stalls low.
The out-of-order windows of hundreds of instructions are also insufficient
to hold all instructions needed in order to maintain a high number of parallel memory requests,
which is necessary to hide long latency accesses.

The two main problems are caused by either branch mispredictions that
make the effective instruction window too small, or by overflowing the
instruction window when there are too many instructions between memory
references.  Since a pending load prevents all following instructions from retiring in-order, if
the instruction window resources cannot hold new instructions, no
concurrent loads can be issued.  A
vicious cycle forms where low ILP causes low MLP when long dependence
chains and mispredicted branches do not generate enough parallel
memory requests.  In turn, low MLP causes low effective ILP when
mispredicted branch outcomes depend on long latency memory references.

Context switching using high number of hardware threads to hide DRAM
latency was explored in Tera~\cite{alverson90ics-tera}.  Today's
commercial CPUs have vestigial simultaneous multithreading support,
e.g., 2-way SMT on Intel CPUs.  OS thread context switching is unusable as it is 50
times more expensive than a DRAM miss.
We therefore go back to 1950s coroutines~\cite{newell56-IPL} for low latency
software context switching in order to hide variable memory latency
efficiently.

We introduce a simple Instruction and Memory Level
Parallelism (\IMLPnsp) programming model based on concurrent tasks executing
as coroutines.  Coroutines yield execution at annotated long-latency
operations, e.g., memory accesses, long dependence chains, or
unpredictable branches.  Our DSL \CIMPLEsp (Coroutines for
Instruction and Memory Parallel Language Extensions)
separates program
logic from programmer hints and scheduling optimizations.  \Cimplesp allows exploration of
task scheduling and techniques such as buffering, vectorization, pipelining, and prefetching, 
supported in our compiler \textbf{\Cimplesp} for C++.

Prior compilers have been unable to uncover many opportunities for parallel
memory requests.  Critical long latency operations are hidden in deeply
nested functions, as modularity is favored by good software
engineering practices.  Aggressive inlining to expose parallel
execution opportunities would largely increase code cache pressure, which
would interact poorly with out-of-order cores. Compiler-assisted
techniques depend on
prefetching~\cite{mowry92asplos-prefetching,lee12taco-prefetch}, e.g.,
fixed look-ahead prefetching in a loop nest.  Manual techniques for
indirect access prefetching have been found effective for the tight
loops of database hash-join operations
~\cite{chen04icde-join-prefetch,kocberber15vldb-amac,menon17vldb-fusion,psaropoulos17vldb-interleaving} -
a long sequence of index lookups can be handled in batches (\textit{static scheduling})~\cite{chen04icde-join-prefetch,menon17vldb-fusion},
or refilled dynamically (\textit{dynamic scheduling})~\cite{kocberber15vldb-amac,psaropoulos17vldb-interleaving}.
Since the optimal style may be data type and data distribution dependent,
Cimple allows generation of both scheduling styles, additional code-generation optimizations, as well as better optimized schedulers.

High performance database query
engines~\cite{neumann11vldb-hyperquery,diaconu13sigmod-hekaton,kornacker15cidr-impala,menon17vldb-fusion}
use Just-In-Time (JIT) compilation to remove virtual function call overheads
and take advantage of attendant inlining opportunities.  For example,
in Impala, an open source variant of Google's F1~\cite{shute12sigmod-f1},
query generation uses both
dynamically compiled C++ text and LLVM Instruction Representation
(IR)~\cite{cloudera13impala-codegen}.  Cimple offers much higher performance with lower complexity
than using an LLVM IR builder: Cimple's Abstract Syntax Tree
(AST) builder is close to C++ (and allows verbatim C++ statements).  Most
importantly, low level optimizations keep working on one item at a
time, while Cimple kernels operate on many items in parallel.

We compare \Cimplesp performance against core in-memory database C++ index implementations:
binary search in a sorted array, a binary tree index lookup, a skip list lookup and traversal, and unordered hash table index.
As shown on \figref{speedup-search-intro}, 
we achieve 2.5$\times$ peak throughput on a multi-core system, and on a single-thread  --- 6.4$\times$ higher performance.

The rest of this paper is organized as follows:
In \secref{example}, we walk through an end-to-end use case.
In \secref{hw_back},
we explore peak ILP and MLP capabilities of modern hardware.
We introduce the \IMLP programming model and \CIMPLEsp compiler and
runtime library design in \secref{design}, with more details of the \Cimplesp DSL in \secref{syntax},
and implementation details of \CIMPLEsp transformations in
\secref{implementation}.  We demonstrate expressiveness
by building a template library of core indexes in
\secref{applications}, and performance -- in \secref{evaluation}. \secref{related} surveys related work and \secref{conclusion} concludes.  

\begin{figure}[t]
 \centering
 \includegraphics[width=1.1\columnwidth]
{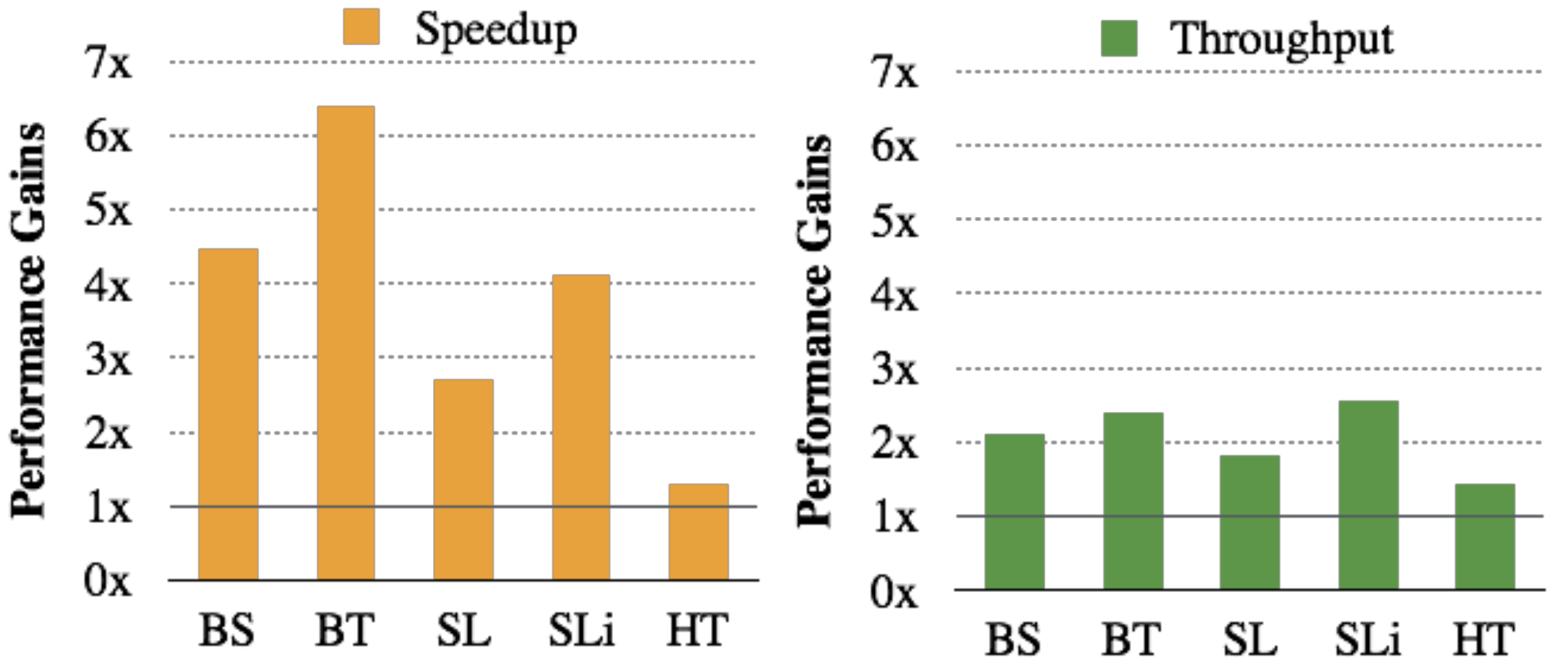}
 \caption{Speedup on a single thread and throughput gains on a full system (24 cores / 48 SMT ~\cite{ark-e5-v3-salike}).
\textbf{B}inary \textbf{S}earch,
\textbf{B}inary \textbf{T}ree,
\textbf{S}kip \textbf{L}ist,
\textbf{S}kip \textbf{L}ist \textbf{i}terator,
and \textbf{H}ash \textbf{T}able. 
}
 \figlabel{speedup-search-intro}
\end{figure}

\section{Example}
\label{sec:example}

We next present an example that highlights how the Cimple language and
runtime system work together to efficiently expose available memory-level
parallelism on current hardware. We use a classic iterative binary search tree lookup,
which executes a while loop to traverse a binary tree in the direction of
\texttt{key} until a match is found. It
returns the node that contains the corresponding key/value
pair, or \texttt{nil}.

\subsection{Binary Search Tree Lookup in Cimple}
\label{sec:examplelookup}

\lstref{bstcimple} presents the Cimple code for the example
computation.  The code identifies the name of the operation
(\texttt{BST\_find}), the result type (\texttt{node*}), and the two
arguments (\texttt{n}, the current node as the computation walks
down the tree, and \texttt{key}, the key to lookup in the tree). 

\begin{flushleft}
\vspace{-8pt}

\begin{listing}[h]
  \centering
\begin{minted}[gobble=4,linenos,escapeinside=~~]{c}
~\textcolor{black}{auto}~ c = Coroutine(BST_find);
c.Result(node*).
Arg(node*, n).
Arg(KeyType, key).
Body().
  While(n).Do(
    ~{\textit{Prefetch(n)}.\textcolor{red}{\textbf{Yield}}().}~
    ~\textbf{If( n->key == key ).}~
    Then( Return(n) ).
    Stmt( n = n->child[~\textbf{n->key < key}~]; )
  ).
  Return(n);
\end{minted}

\caption{Binary Search Tree Lookup in Cimple.}
\label{lst:bstcimple}

\end{listing}

\end{flushleft}

In this code, there is one potentially expensive memory operation,
specifically the first access to \texttt{n->key} in the if condition
that checks to see if the key at the current node \texttt{n} matches
the lookup \texttt{key}.  Once the cache line containing this value has
been fetched into the L1 data cache, subsequent accesses to \texttt{n->key} and \texttt{n->child} are
accessed quickly. The Cimple code issues a prefetch,
then yields to other lookup operations on the same thread.

\begin{flushleft}
\begin{listing}
  \centering

\begin{minted}[xleftmargin=10pt,gobble=4,fontsize=\small,linenos,escapeinside=~~]{c}
struct Coroutine_BST_Find {
  node* n;
  KeyType key;
  node* _result;
  int _state = 0;
  enum {_Finished = 2};

  bool Step() {
    switch(_state) {
    case 0:
      while(n) {
  ~\textit{      prefetch(n)}~;
        _state = 1;
        return false;
    case 1:
        if(n->key == key) {
          _result = n;
          _state = _Finished;
          return true;
        }
        n = n->child[n->key < key];
      } // while
      _result = n;
      _state = _Finished;
      return true;
    case _Finished:
      return true;
  }}};
    \end{minted}
\caption{Generated Cimple coroutine for \texttt{BST\_find}. }
\label{lst:bstcoro}

\end{listing}

\end{flushleft}

\lstref{bstcoro} presents the coroutine that our Cimple compiler
(automatically) generates for the code in
\lstref{bstcimple}. Each coroutine is implemented as a C++
struct that stores the required state of the lookup computation and
contains the generated code that implements the lookup. The
computation state contains the \texttt{key} and current node \texttt{n}
as well as automatically generated internal state variables
\texttt{\_result} and \texttt{\_state}. Here after the Cimple compiler has
decomposed the lookup computation into individual steps,
the computation can be in one of three states:

{\bf Before Node:} In this state the lookup is ready to check if
  the current node \texttt{n} contains \texttt{key}. However, the
  required access to \texttt{n->key} may be expensive. The step
  therefore issues a prefetch on \texttt{n} and returns back to the
  scheduler. To expose additional memory level parallelism and hide
  the latency of the expensive memory lookup, the scheduler will
  proceed on to multiplex steps from other lookup computations onto
  the scheduler thread.

{\bf At Node:} Eventually the scheduler schedules the next step
  in the computation. In this step, the prefetch has (typically)
  completed and \texttt{n} is now resident in the L1 cache. The
  computation checks to see if it has found the node containing the
  key. If so, the lookup is complete, the coroutine stores the found
  node in \texttt{\_result}, and switches to the Finished
  state. Otherwise, the coroutine takes another step left or right
  down the search tree, executes the next iteration of the while loop
  to issue the prefetch for left or right node, and then returns back
  to the scheduler.

{\bf Finished:} Used only by schedulers that
execute a batch of coroutines that require different number of steps.

\subsection{Request Parallelism}  

Cimple converts available request level parallelism (RLP) into
memory-level parallelism (MLP) by
exposing a queue of incoming requests to index routines, instead of
queuing or batching in the network stack~\cite{li15isca-mica}.
Our example workload is inspired by modern Internet
servers~\cite{armstrong13sigmod-linkbench,bronson13atc-tao,rocksdb17}
that process a high volume of aggregated user requests.
Even though the majority of requests are for key lookups, support for
range queries requires an ordered dictionary, such as a binary search
tree or a skip list.  Here each worker thread is given a stream of
independent key lookup requests.

A coroutine \textit{scheduler} implements a lightweight, single-threaded
queue of in-flight partially completed request computations (e.g., BST lookups).
The scheduler multiplexes the computations onto its thread at
the granularity of steps.
The queue
stores the state of each partially completed computation and switches
between states to multiplex the multiple computations. The Cimple implementation breaks each computation into a
sequence of steps. Ideally, each step performs a sequence of local
computations, followed by a prefetch or expensive memory access
(e.g., an access that is typically satisfied out of the DRAM), then a yield. 

Note we never wait for events, since loads are not
\textit{informing}~\cite{horowitz96isca-informing}.
We simply avoid reading values that might stall.
This is the fundamental difference between Cimple and heavy-weight event-driven I/O schedulers.
We also avoid non-predictable branches when resuming coroutine stages.

We maintain a pipeline of outstanding requests that covers the maximum
memory latency.  The scheduler queue has a fixed number of entries,
e.g., \verb|~|50 is large enough to saturate the memory level
parallelism available on current hardware platforms.  The scheduler executes
one step of all of queued computations.  A queue refill is requested
either when all lookups in a batch complete
(\textit{static scheduling}~\cite{chen04icde-join-prefetch}),
or as soon as any lookup in a batch has completed (\textit{dynamic scheduling}~\cite{kocberber15vldb-amac}).
The scheduler then returns
back to check for and enqueue any newly arrived requests. In this way
the scheduler continuously fills the queue to maximize the exploited
memory level parallelism.

\subsection{Cimple Execution On Modern Computing Platform}
For large binary search trees, the aggregated lookup computation is
memory bound.  Its performance is therefore determined by the
sustained rate at which it can generate the memory requests required
to fetch the nodes stored in, e.g., DRAM or other remote memory.  Our
target class of modern microprocessors supports nonblocking cache
misses, with up to ten outstanding cache misses in flight per core at any given
time. The goal is therefore to maximize the number of outstanding
cache misses in flight, in this computation by
executing expensive memory accesses from different computations in
parallel.

Here is how the example Cimple program works towards this goal.  By
breaking the tree traversals into steps, and using the Cimple
coroutine mechanism to quickly switch between the lookup computation
steps, the computation is designed to continuously generate memory
requests (by issuing prefetch operations from coroutined lookups). 
This execution strategy is designed to generate an instruction stream
that contains sequences of fast, cache-local instructions (from 
both the application and the Cimple coroutine scheduler) interspersed
with prefetch operations. While this approach has instruction overhead
(from the Cimple coroutine scheduler), the instructions execute quickly
to expose the available MLP in this example.

\subsection{Performance Comparison} 
\begin{figure}
 \centering
 \includegraphics[width=.9\columnwidth]
{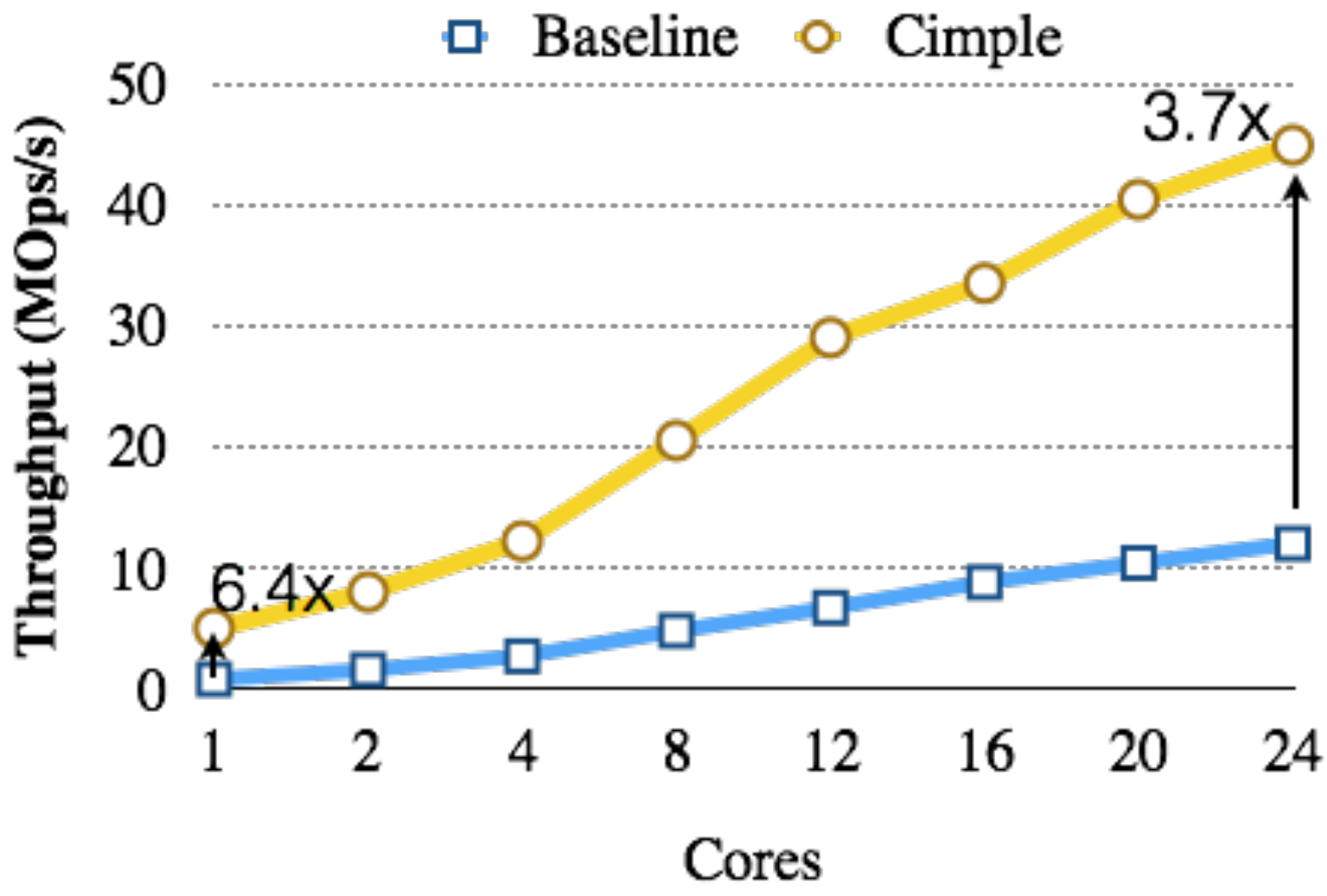}
 \caption{Throughput improvements for lookup in a partitioned binary search tree index (1GB per thread).
}
 \label{fig:lookup-salike}
\end{figure}

We compare the performance of the Cimple binary tree lookup with 
the performance of a baseline binary tree lookup algorithm.
The workload is a partitioned tree search in which each thread is given 
a stream of lookups to perform. The Cimple implementation interleaves multiple
lookups on each thread, while the baseline executes
the lookups sequentially. 
We use a 24 core Intel Haswell machine with 2 hyperthreads per core (see \secref{hwcfg}).

\figref{lookup-salike} presents the results. The X axis is the number
of cores executing the computation, with each core executing a single
lookup thread. The Y axis presents the number of lookups in millions
of operations per second. On one thread, the Cimple computation performs
6.4 times as many lookups per second as the baseline computation. This
is true even though 1) due to coroutine scheduling overhead,
the Cimple computation executes many more instructions than the baseline
computation and 2) in theory, the baseline computation has as much 
memory parallelism across all requests as the Cimple computation (but the baseline MLP is unavailable
to the processor because it is separated within the instruction stream
by the long sequential lookups).

The performance of both computations increases up to 24 cores, with the
Cimple implementation performing 3.7 times as many lookups per second
as the baseline implementation (the difference narrows because the 
memory and coherence systems become increasingly loaded as the number of cores
increases). Our machine supports two hyperthreads per core. Increasing
the number of threads from 24 to 48 requires placing two threads on
at least some of the cores. With this placement, the Cimple threads
start interfering and performance decreases. The performance of the baseline
computation increases (slowly) between 24 and 48 threads. Nevertheless, 
the best Cimple computation (on 24 threads) still performs 2.4 times
as many operations per second as the best baseline computation
(on 48 threads).

\subsection{Three Key Techniques to Improve MLP and ILP } 
\label{sec:threekeys}

The example on \lstref{bstcimple} illustrates the three essential
techniques for achieving good performance with Cimple on current
hardware.  The first and most important is to identify
independent requests and allow parallelism across them by breaking up execution at \keyword{Yield}
statements (line 7).  The second is to enable longer computation chains
between memory requests via explicit software prefetching
\keyword{Prefetch}. The third is to eliminate unpredictable branches --- by replacing a
control dependence (if) with an address generation dependence (line
10). Otherwise branch mispredictions would also discard unrelated (correct) subsequent
coroutines, since hardware speculative execution is designed to
capture the control flow of only one sequential instruction stream.

\section{Hardware Background}
\label{sec:hw_back}

We now examine the hardware mechanisms for handling cache misses and
memory level parallelism in DRAM and CPUs.  The achievable MLP is further
limited by the size of the buffers connecting memory hierarchy layers.

\subsection{DRAM Parallelism}
\label{sec:drampar}
The two main MLP limiters are the number of DRAM banks, and the size of pending request queues.

Before a DRAM read or write can occur, a DRAM row -- typically 8--16KB
of data -- must be destructively loaded (\textit{activated}) into an
internal buffer for its contents to be accessed.
Consecutive read or write requests to the same row are limited only by
DRAM channel bandwidth, thus sequential accesses take advantage of
spatial locality in row accesses. In contrast, random accesses must
find independent request streams to hide the high latency of a row
cycle of different banks (\textit{DRAM page misses}), or worse -- the
additional latency of accessing rows of the same bank (\textit{DRAM
  page conflicts}).
The typical maximum of simultaneously open banks on
a DDR4 server is 16 banks$\times$2 ranks$\times$(4---6) memory channels.
The memory controllers track more pending requests in large queues, e.g., 48+ cache lines per memory channel
~\cite{esmer15micro-ivb}.

While DRAM latency has stagnated, higher DRAM bandwidth and more
memory controllers have kept up with providing high per-core memory
bandwidth.  The share of total bandwidth for cores on current Intel
Xeon servers is 4--6 GB/s.  Although DDR4 latency is
$\sim$\SI{50}{ns}, additional interconnect, cache coherence, and queuing
delays add to total memory latency of 80~ns--200~ns.

\subsection{Cache Miss Handling}
\label{sec:mshrs}
The primary MLP limit for single threaded execution is the number of
Miss Status Holding Registers (MSHR)~\cite{kroft81isca-mshr}, which are the
hardware structures that track cache lines with outstanding cache misses.
Modern processors typically have 6--10 L1 cache MSHRs: since a
content-associative search is expensive in area and power, the number
of MSHRs is hard to scale~\cite{tuck06micro-mshr}.  Intel's Haswell
microarchitecture uses 10 L1 MSHRs (Line Fill Buffers) for handling
outstanding L1 misses~\cite{intel17opt}.  
The 16 L2 MSHRs limit overall random and prefetchable sequential traffic.

For current software with low MLP, the MSHRs are hardly a bottleneck.
Hardware prefetching and speculative instructions (after branch
prediction) are important hardware techniques that put to use the rest
of the MSHRs.  Hardware prefetching is effective for sequential
access -- in that case a few MSHRs are sufficient to hide the access
latency to the next level in the memory hierarchy.  When hardware
prefetches are wrong, or when mispredicted branches never need the
speculatively-loaded cache lines, these techniques are wasting memory
bandwidth and power.

By Little's law, the achievable bandwidth equals the number of MSHR
entries divided by the average memory latency. 
Applications that are stalling on memory but do not saturate the
memory bandwidth are typically considered ``latency bound''.  More often
than not, however, the real bottleneck is in the other term of
Little's law - a very low queue occupancy due to low application MLP.
The effective MLP of several graph frameworks is estimated in~\cite{beamer15iiswc-ivybridge}.

\subsection{Software Prefetching}
Using software prefetch instructions allows higher MLP than regular loads.
The instruction reorder buffer, or any resource held up by non-retired
instructions may become the limiting factor: 192-entry reorder buffer, 168 registers, 72-entry load and
42-entry store buffers on Haswell~\cite{intel17opt}.  
These resources are plentiful
when running inefficiently one memory request at a time.
Dividing the core resources over 10 parallel memory requests,
however, means that each regular load can be accompanied by
at most 19 $\micro$ops using at most 16 registers, 7 loads and 4 memory stores.

Prefetch instructions free up the instruction window as they retire
once the physical address mapping is known, e.g., either after a TLB hit, 
or after a page walk on a TLB miss.  As soon as the virtual to
physical translation has completed, an L2 memory reference using
the physical address can be initiated.
On current Intel microarchitectures the PREFETCH\textit{h} family of instructions always prefetch into the L1 cache. Software prefetches are primarily limited by
the number of L1 MSHR entries.
Maintaining a longer queue of in-flight requests (limited by load
buffers), however, helps to ensure that TLB translations of the
following prefetches are ready as soon as an MSHR entry is
available.  If hardware performance counters show that dependent
loads miss both the L1 cache and MSHRs then prefetches are too
early; if loads hit MSHRs instead of L1 then prefetches are too late.

\subsection{Branch Misprediction Handling}
Highly mispredicted branches are detrimental to speculative execution,
especially when a burst of branch mispredictions results in a short
instruction window.  Mispredicted branches that depend on long latency loads also incur a high
speculation penalty.  Instruction profiling with hardware performance
counters can be used to precisely pinpoint such critical branches.
The most portable solution for avoiding branch misprediction is to use
data or address dependence instead of control dependence.  While no
further execution of dependent instructions is possible,
independent work items can still be serviced.

Most instruction set architectures (ISAs) also support conditional move
instructions (\texttt{cmov} on x86) (or \texttt{csel} on ARM), as simple cases of instruction predication.  Automatic predication is
also available in simple cases on IBM Power7~
\cite{sinharoy11ibm-power7-autopredication} where unpredictable
branches that jump over a single integer or memory instructions are
converted to a predicated conditional selection.  The ternary select
operator in C (\texttt{?:}) is often lowered to conditional move instructions,
however, use of assembly routines is unfortunately required to ensure that
mispredictable branches are not emitted instead.

\section{Design Overview}
\label{sec:design}

The \CIMPLEsp compiler and runtime library are used via an embedded
DSL similar to Halide~\cite{jrk13pldi-halide}, which separates the basic logic
from scheduling hints to guide transformations.  Similarly we build
an Abstract Syntax Tree (AST) directly from succinct C++ code.
Unlike Halide's expression pipelines, which have no control flow,
\Cimplesp treats expressions as opaque AST blocks and exposes
conventional control flow primitives to enable our transformations.
\secref{syntax} describes our \Cimplesp syntax in more detail.

Coroutines are simply routines that can be interleaved with other
coroutines.  Programmers annotate long-latency operations, e.g.,
memory accesses, or unpredictable branches. A \keyword{Yield} statement marks
the suspension points where another coroutine should run.
Dynamic coroutines are emitted as routines that can be resumed at the
suspension points, and a \texttt{struct} tracking all live variables.

\lstref{bstcimple} showed lookup in a traditional Binary Search Tree
written in \Cimplensp.  A coroutine without any Yield statements is simply a routine,
e.g., a plain C++ routine can be emitted 
to handle small-sized data structures, or if Yield directives are disabled. 
The bottleneck in \lstref{bstcimple} is the expensive pointer
dereference on line 8.  Yet, prefetching is futile unless we context switch
to another coroutine.  \lstref{bstcoro} showed a portable~\cite{duff88device} unoptimized coroutine for a dynamic
coroutine scheduler (\secref{dyncoro}).

\subsection{Target Language Encapsulation}
\CIMPLEsp emits coroutines that can be included directly in the
translation unit of the original routines.  
Our primary DSL target language is C++.
All types and expressions are opaque;
statements include opaque raw strings in the syntax of the target language, e.g.,
native C++.

\subsection{\Cimplesp Programming Model}
A coroutine yields execution to peer coroutines only at \textbf{Yield} suspension
points.

Expensive memory operations should be tagged with
\textbf{Load} and \textbf{Store} statements (which may yield according to a scheduling policy), or with an
explicit \Syntax{Prefetch} directive (see \secref{memory}).  Loads and stores that hit caches can simply
use opaque native expressions.

\Syntax{If}/\Syntax{Switch} or \Syntax{While}/\Syntax{DoWhile} statements should be used
primarily to encapsulate mispredicted branches.  Most other
control-flow statements can use native selection and iteration statements.

A coroutine is invoked using a coroutine scheduler.
Regular routines are called from coroutines as usual in statements and expressions.
Coroutines are called from inside a coroutine with a \textbf{Call}.

\subsection{Scheduling Hints}
Scheduling hints guide transformations and can be added as annotations to the corresponding memory access or control statements.
The example in \lstref{bstcimple} showed
how a single source file handles four largely orthogonal concerns.
First, the program structure is described in \Cimplensp,
e.g., \SyntaxTail{While}.  Second, optional inline scheduling directives are specified, e.g.,
  \SyntaxTail{\mbox{Yield}}.
Third, out-of-line scheduling can access AST node handles in C++, e.g., \verb|auto c|.
Finally, all target types and expressions are used unmodified, e.g.,
\verb|list*|.

\subsection{Parallel Execution Model}
Cimple coroutines are interleaved only on the creating thread to
maintain a simple programming model, and to enable efficient
scheduling (\secref{calls}).
\IMLPsp composes well with thread and task
parallelism~\cite{blumofe95cilk,openmp-45}.  Instead of running to
completion just a single task, a fork-join scheduler can execute
concurrently multiple coroutines.
The embedded DSL approach allows easy 
integration with loop and task
parallelism extensions, e.g., \verb|#pragma| extensions integrated with OpenMP
~\cite{openmp-45} or Cilk~\cite{blumofe95cilk,leiserson10jsc-cilkp}.
\section{{\textsc \Cimplesp} Syntax and Semantics}
{\large }
\label{sec:syntax}

An original C++ program is easily mapped to the conventional control
flow primitives in \Cimplensp.  \tabref{stmt-summary} summarizes our
statement syntax and highlights in bold the unconventional directives.

\subsection{Coroutine Return}
\label{sec:return}
A coroutine may suspend and resume its execution at specified with \SyntaxFirst{Yield} suspension points,
typically waiting on address generation, data, or branch resolution.  Programmers
must ensure that coroutines are reentrant.

\Syntax{Return} stores a coroutine's result, but does not return to
the caller, instead it may resume the next runnable coroutine.  \Syntax{Result}
defines the coroutine result type, or \texttt{void}.

\subsection{Variable Declarations}
\label{sec:variable}

The accessible variables at all coroutine suspension points form its
context.  A target routine's internal variables need to be declared
only when their use-def chains cross a yield suspension point.  A
\SyntaxFirst{Variable} can be declared at any point in a block and is
presumed to be live until the block end.  \textbf{Arg}uments to a
coroutine and its \textbf{Result} are Variables even when internal
uses do not cross suspension points.  Shared arguments among
coroutines using the same scheduler can be marked \textbf{SharedArg} to reduce register pressure.

References in C++ allow
variables to be accessed directly inside opaque expressions, e.g.:

\lstset{emph={Arg,Variable},emphstyle=\bfseries}
\lstinline|    Arg(int, n).Variable(int, x, {n*2})| \\ %
For C \SyntaxTail{Variable} accesses must use a macro: \verb|Var(a)|.  We do not analyze variable uses in
opaque expressions, but judicious block placements can minimize a
variable's scope.

\begin{table}[t]
\footnotesize
 \centering
 \begin{tabular}{ll}
\textbf{Statement} & \textbf{Section} \\
 \texttt{Arg}, \texttt{SharedArg}, \texttt{Result}, \texttt{Variable} & \secref{variable} \\
 \texttt{If}, \texttt{Switch}, \texttt{Break} & \secref{ifswitch} \\
 \texttt{While}, \texttt{DoWhile}, \texttt{Continue}  & \secref{while} \\
 \texttt{Call} & \secref{calls} \\
 \texttt{Return}, \textbf{\texttt{Yield}} & \secref{return} \\
 \texttt{Load}, \texttt{Store}, \texttt{Assign}, \textbf{\texttt{Prefetch}} & \secref{memory} \\
  \end{tabular}
  \caption{\Cimplesp Statements.}
  \tablabel{stmt-summary}
\end{table}

\subsection{Block Statement}
\label{sec:block}

A block statement encapsulates a group of statements and declarations. Convenience macros wrap the verbose Pascal-like \texttt{Begin} and \texttt{End} AST nodes, e.g., we always open a block for the \textbf{Then}/\textbf{Else} cases in \texttt{If}, \textbf{Do} in \texttt{While}, and
\textbf{Body} for the function body block.

\subsection{Opaque Statements and Expressions}
\label{sec:opaque}

Types and expressions used in \Cimplesp statements are strings
passed to the target compiler.  Opaque statements are created from
string literals, though convenient preprocessor macros or C++11 \textit{raw strings}
allow clean multi-line strings and unmodified code wrapping in a
\texttt{.cimple.cpp} source file, e.g.: \\ %
\lstset{language=[ISO]C++,showspaces=false}
\vspace{-14pt}
\begin{minted}[xleftmargin=10pt,gobble=4,fontsize=\footnotesize,escapeinside=~~]{c}
  << R""( // Murmur3::fmix32
       h ^= h >> 16; h *= 0x85ebca6b;
       h ^= h >> 13; h *= 0xc2b2ae35;
       h ^= h >> 16;
     )""
\end{minted}

\subsection{Selection Statements}
\label{sec:ifswitch}
\textbf{\texttt{If}} and \textbf{\texttt{Switch}} selection
statements, can be used for more effective if-conversion to avoid
mispredicted branches.  If/Switch give more control over branch-free
if-conversion, for conventional 2-way branch and case selection.

Well-predicted branches do not need to be exposed, and can simply use
native \texttt{if/switch} in opaque statements.  Opaque conditional
expressions (\verb|?:|) and standard if-conversion, that converts
branches into conditional moves, are effective when only data content
is impacted.  Traditional predicated execution and conditional moves
are less effective when address dependencies need to be hidden,
especially for store addresses.  Predicated execution also 
inefficiently duplicates both sides of a branch.

A \Syntax{Switch} must also be used instead of a \texttt{switch} when
a case has a suspension point, see \secref{dyncoro}.

\subsection{Iteration Statements}
\label{sec:while}

\texttt{While} and \texttt{DoWhile} iteration statements are exposed
to \Cimplesp when there are internal suspension points to enable
optimizations. Conventional \texttt{Continue} and \texttt{Break} respectively skip
the rest of the body of an iteration statement, or terminate the body of
the innermost iteration or \texttt{Switch} statement.

\subsection{Informed Memory Operations}
\label{sec:memory}

\Syntax{Load} and \Syntax{Store} statements mark expensive memory
operations that may be processed optimally with one or more internal
suspension points.  \Syntax{Prefetch} explicitly requires that one or more
independent prefetches are issued before yielding.
\Syntax{Assign} can mark explicitly other assignments that are expected to be operating on cached data.

\subsection{Coroutine Calls}
\label{sec:calls}

Tail-recursive \SyntaxFirst{Call} resumes execution to the initial
state of a coroutine.  Regular function calls
can be used in all expressions, and are inlined or called as routines as usual.
A Return calling a {\tt void} coroutine is also allowed, as in C++, for explicit tail-recursion.

\section{DSL Compiler and Runtime Library}
\label{sec:implementation}

The DSL allows exploration of multiple coroutine code generation
variants and combinations of data layout, code structure, and runtime
schedulers.
We use two main code generation strategies for handling a \textit{stage}
(the code sequence between two \texttt{Yield} statements, or function entry/exit): 
\textit{static} where a stage becomes a \texttt{for} loop body, and \textit{dynamic} where a stage
forms a \texttt{switch} case body.  The \texttt{Yield} directive marking the boundary of
a coroutine stage can select the schedule explicitly.

We first discuss the context of a single coroutine,
and storage formats for tracking active and pending coroutines.
Then we discuss how these are used in runtime schedulers that create, execute, and retire coroutines.

\subsection{Coroutine Context}
\label{sec:corocontext}

A coroutine's closure includes all private arguments and variables of
a coroutine.  Shared arguments between instances are stored only once per
scheduler and reduce register pressure.  Additional variables are optionally stored in the context depending on the
code generation choices:
a Finite State Machine \texttt{state} is used for dynamic scheduling on Yield;
a \texttt{result} value (of user-defined type) holds the final result; 
a \texttt{cond}ition -- when If yields before avoid making decisions on hard to resolve branches;
an \texttt{addr}ess (or index) -- when Load/Store yield before using a hard to resolve address.

\begin{minted}[xleftmargin=10pt,gobble=4,fontsize=\small,escapeinside=~~]{c}
struct BST::find__Context_AoS {
  node* n;     // Arg
  KeyType key; // Arg
  ~\textcolor{black}{int}~   _state;  // for dynamic Yield
  ~\textcolor{black}{node*}~ _result; // for Return
  ~\textcolor{black}{bool}~  _cond;   // for If
  ~\textcolor{black}{void*}~ _addr;   // for Load/Store
\end{minted}

\paragraph{Vectorization-friendly Context Layout}
\label{sec:vectorization}
The primary distinctive design choice of \Cimplesp is that we need to
run multiple coroutines in parallel, e.g., typically tens.
For homogeneous coroutines we choose between Struct-of-Array (SoA), Array-of-Struct (AoS), and Array-of-Struct-of-Array (AoSoA) layouts.
Variable accesses are insulated from these changes via convenient C++ references.

\subsubsection{Static Fused Coroutine}
\label{sec:staticfuse}

Homogeneous coroutines that are at the same stage of execution
can be explicitly unrolled, or simply emitted as a loop.
The target compiler has full visibility inside any inlined functions
to decide how to spill registers, unroll, unroll-and-jam, or vectorize.
An example of SIMD vectorization of a hash function
(\lstref{htlprobe} in \secref{hashjoin}) is shown on \lstref{corostatic}.
The hash finalization function called on line 5
has a long dependence chain (shown inlined earlier in \secref{opaque}).
C++ references to variables stored in SoA layout, shown on lines 3--4 and 9--10, allow the opaque statements to access all Variables as usual.

\begin{flushleft}

\begin{listing}[H]
  \centering
\begin{minted}[xleftmargin=10pt,gobble=4,fontsize=\small,linenos,escapeinside=~~]{c}
  bool SuperStep() { 
    for(int _i = 0; _i < _Width ; _i++) {
      ~{\textit{KeyType& k = \_soa\_k[\_i];}}~
      ~{\textit{HashType& hash = \_soa\_hash[\_i];}}~
        hash = Murmur3::fmix(k);
        hash &= mask;
    }
    for(int _i = 0; _i < _Width ; _i++) {
      KeyType& k = _soa_k[_i];
      HashType& hash = _soa_hash[_i];
        prefetch(&ht[hash]);
    }
\end{minted}
\caption{Stages of a Static Coroutine for \lstref{htlprobe}.}
\label{lst:corostatic}

\end{listing}
\end{flushleft}

Exposing loop vectorization across strands offers an opportunity for
performance gains.  Since we commonly interleave multiple instances of
the same coroutine, we can fuse replicas of the basic blocks of the
same stage working on different contexts, or stitch different stages
of the same coroutine, or even different coroutines.  These are
similar to unroll-and-jam, software pipelining, or function
stitching~\cite{gopal10stitching}.  Stage fusion benefits
from exposing more vectorization opportunities, reducing scheduling
overhead, and/or improving ILP.

Basic block vectorization, e.g., SLP~\cite{larsen00pldi-slp}, can be
improved by better Variable layout when contexts are stored in array of
struct (AoS) format. 

\subsubsection{Dynamic Coroutine}
\label{sec:dyncoro}

Coroutines may be resumed multiple times unlike one-shot
continuations.  Typical data structure traversals may require
coroutines to be suspended and resumed between one and tens of times.

We showed on \lstref{bstcoro} the basic structure of a \texttt{switch} based coroutine that uses
a ``Duff's device''~\cite{duff88device} state machine tracking.  This method takes advantage of the loose syntax of a
\texttt{switch} statement in ANSI C.  Surprisingly to some,
\texttt{case} labels can be interleaved with other control flow, e.g.,
\texttt{while} loops or \texttt{if}.  Only enclosed \texttt{switch}
statements can not have a suspension point. Mechanical addition of case labels within the existing control flow is
appealing for automatic code generation: we can decorate the original
control flow graph with jump labels at the coroutine suspension points and add a top
level \texttt{switch} statement.

This standard C syntax allows good portability
across compilers.  However, the reliance of a \texttt{switch} and labels
precludes several optimization opportunities.
The alternatives include relying on computed \texttt{goto} (a
\texttt{gcc} extension), indirect jumps in assembly, or method function
pointers as a standard-compliant implementation for C++.  The first two are less portable, while the latter results in code
duplication when resuming in the middle of a loop.

Short-lived coroutines suffer from branch mispredictions on stage
selection, and using a \texttt{switch} statement today leaves to compiler
optimizations, preferably profile guided, to decide between using a
jump table, a branch tree, or a sequence of branches sorted by
frequency.  Unlike threaded interpreters, which benefit from correlated pairs of
bytecode, ~\cite{ertl03pldi-threaded,rohou15cgo-interp}, the potential
correlation benefits from threading coroutines come from burstiness across requests.
An additional optimization outside of the traditional
single coroutine optimization space is to group across coroutines
branches with the same outcome, e.g., executing the same stage.

\subsection{Coroutine Schedulers}
\label{sec:schedulers}

We discuss the salient parameters of coroutine runtime scheduler flavors, and
their storage and execution constraints.  
We target under 100 active
coroutines (\textit{Width}) with under 100B state each to stay L1-cache resident.  Below is a
typical use of a simple coroutine scheduler (for \lstref{csliter}):
\begin{minted}[xleftmargin=10pt,gobble=4,fontsize=\footnotesize,linenos,escapeinside=~~]{c}
template<~\textcolor{black}{int}~ ~\textcolor{purple}{\bf Width}~ = ~\textcolor{black}{\bf 48}~>
~\textcolor{black}{void}~ SkipListIterator_Worker(~\textcolor{black}{size\_t}~* answers,
                      node** iter, ~\textcolor{black}{size\_t}~ len) {
   using Next = CoroutineState_SkipList_next_limit;
   SimplestScheduler<Width, Next>(len,
      [&](Next* cs, ~\textcolor{black}{size\_t}~ i) {
            *cs = Next(&answers[i], IterateLimit,
                       iter[i]);
   });
}
\end{minted}

\paragraph{Static Batch Scheduler}
Tasks are prepared in batches similar to manual \textit{group prefetching}~\cite{chen04icde-join-prefetch,menon17vldb-fusion}.
Storage is either static AoS, or in SoA format to support vectorization.
Scaling to larger batches is less effective if tasks have variable completion time, e.g., on a binary search tree.
Idle slots in the scheduler queue result in low effective MLP.

\paragraph{Dynamic Refill Scheduler}
Tasks are added one by one, and refilled as soon as a task completes,
similar to the manual approach in AMAC~\cite{kocberber15vldb-amac}.  
Storage is in static or dynamic-width AoS.
Further optimizations are needed to reduce branch mispredictions to improve effective MLP.

\paragraph{Hybrid Vectorized Dynamic Scheduler}
Hybrid across stages, where the first stages of a computation can use a static scheduler,
but following stages use a dynamic scheduler while accessing the SoA layout.

\subsubsection{Common Scheduler Interface}
Runtime or user-provided schedulers implement
common APIs for initialization, invoking coroutines, and draining results.
A homogeneous scheduler runs identical coroutines with the same shared arguments.
New tasks can either be pulled via a scheduler callback, or pushed
when available. A pull task with long latency operations or branch
mispredictions, may become itself a bottleneck.
Routines with non-\texttt{void} results can be drained either in-order
or out-of-order.  Interfaces can drain either all previous tasks, or until
a particular task produces its result.

We show in \appref{dynsimplest}
the simplest scheduler and a typical scheduler use.
Using \texttt{push/pull} callback functors as an alternative to enqueue/dequeue initiated by an outside driver,
is shown in \appref{pushpull}.

\section{Applications}
\label{sec:applications}

We study \Cimplensp's expressiveness and performance on core database
data structures and algorithms.  Simple near-textbook
implementations in \Cimplesp ensure correctness, while scheduling directives are used to fine-tune
performance.  We compare \CIMPLEsp C++, against \naive
native C++, and optimized
baselines from recent research.

We start with a classic binary search, which is often the most efficient solution for a read-only dictionary.
For a mutable index, 
in addition to the binary search tree we have shown in \secref{example}, here we show a skip list.
Since both of these data structures support efficient range queries in addition to lookup,
these are the default indices of VoltDB, and RocksDB respectively.  
Finally, we show a hash table as used for database join queries.

\subsection{Array Binary Search}

\begin{flushleft}

\begin{listing}
  \centering
\begin{minted}[xleftmargin=10pt,gobble=4,fontsize=\small,linenos,escapeinside=~~]{c}
  Arg(ResultIndex*, result).
  Arg(KeyType, k).
  Arg(Index, l).
  Arg(Index, r).
  Body().
  While( l != r ).Do(
    Stmts(R""( {
      ~\textcolor{black}{int}~ mid = (l+r)/2;
      ~\textcolor{black}{bool}~ less = (a[mid] < k);
      l = less ? (mid+1) : l;
      r = less ? r : mid;
       } )"").
     Prefetch(&a[(l+r)/2]).Yield()
  ).
  Stmt( *result = l; );
\end{minted}

\caption{Cimple binary search. }

\label{lst:bsearch}
\vspace{-8pt}
\end{listing}

\end{flushleft}

\lstref{bsearch} shows our Cimple implementation.
Current \texttt{clang} compilers use a conditional move for the ternary operators on lines 10--11.
However, it is not possible to guarantee that compilers will not revert to using a branch,
especially when compiling without Profile Guided Optimization.
Programmers can choose to write inline assembly using raw statements for finer control, or use provided helper functions. 

Perversely, a \naive baseline performs better with a mispredicted branch
as observed in ~\cite{khuong17jea-bsearch}, since 
speculative execution is correct 50\% of the time. 
When speculative loads have no address dependencies
hardware aggressively prefetches useless cache lines, as we show in \secref{perfcounters}.

The \Cimplesp version works on multiple independent binary searches 
over the same array. All of our prefetches or memory loads are useful.

\subsection{Skip List}

\begin{figure}[h]
 \centering
 \includegraphics[trim=0cm 15cm 12cm 0cm, clip, width=\columnwidth]
{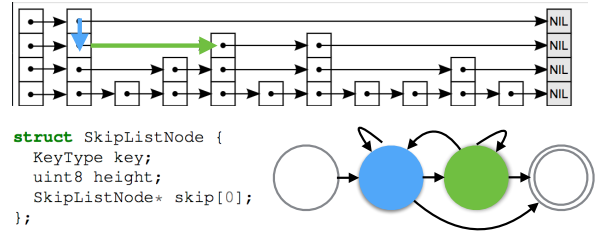}
 \caption{SkipList Layout, and Coroutine State Machine
}
 \label{fig:slfsm}
\end{figure}

\begin{flushleft}
\begin{listing}
  \centering
\begin{minted}[xleftmargin=10pt,gobble=4,fontsize=\small,linenos,escapeinside=~~]{c}
VariableInit(SkipListNode*, n, {}).
VariableInit(uint8, ht, {pred->height}).
While(true).Do(
    While(ht > 0).Do( // down
      Stmt( n = pred->skip[ht - 1]; ).
      Prefetch(n).Yield().
      If(!less(k, n->key)).Then(Break()).
      Stmt( --ht; )
    ).
    If (ht == 0).Then( Return( nullptr )).
    Stmt( --ht; ).
    While (greater(k, n->key)).Do(
      Stmt( pred = n; n = n->skip[ht]; ).
      ~{\textit{Prefetch(n)}.\texttt{Yield}().}~
    ).
    If(!less(k, n->key)).Then(
       Return( n )));
\end{minted}
\caption{Cimple Skip List lookup.}
\label{lst:cslfind}
\vspace{-8pt}
\end{listing}

\end{flushleft}

\begin{flushleft}
\begin{listing}
  \centering
\begin{minted}[xleftmargin=10pt,gobble=4,fontsize=\small,linenos,escapeinside=~~]{c}
    While( limit-- ).Do(
        Prefetch(n).Yield().
        Stmt( n = n->skip[0]; )
    ).
    Prefetch(n).Yield().
    Return( n->key );
\end{minted}
\caption{Cimple Skip List Iteration.}
\label{lst:csliter}
\vspace{-8pt}
\end{listing}

\end{flushleft}

\paragraph{Lookup}
Our skip list baseline is Facebook's \texttt{folly} template library
implementation of \textsc{ConcurrentSkipList}~\cite{herlihy2006provably}.
\figref{slfsm} shows the skip list data structure layout, and the state machine generated for
the code on \lstref{cslfind} which follows \textit{down} and then \textit{right}.
Note that in the \textit{down} direction (line 5) an array of pointers
is explored, therefore speculative execution in the baseline is not blocked by address dependencies;
the \textit{right} direction (line 13) cannot be speculated.

\paragraph{Range Query}
\label{sec:rangequery}
Range queries are the main reason ordered dictionaries are used as default indices.
Skip lists iteration requires constant, but still inefficient, pointer chasing (\lstref{csliter}).
Request level parallelism in range queries is handled similarly to lookup by
interleaving multiple independent queries for both finding the first
node, and for iterating and aggregating over successive nodes.

\subsection{Hash tables}
\label{sec:hashjoin}

\begin{flushleft}

\begin{listing}
  \centering
\begin{minted}[xleftmargin=10pt,gobble=4,fontsize=\small,linenos,escapeinside=~~]{c}
Result(~\textcolor{black}{KeyValue*}~).
Arg(KeyType, k).
Variable(HashType, hash).
Body().
  Stmt ( hash = Murmur3::fmix(k); ).
  Stmt ( hash &= this->size_1; ).Yield().
  Prefetch( &ht[hash] ).Yield()
  << R""(
  while (ht[hash].key != k &&
         ht[hash].key != 0) {
       hash++;
       if (hash == size) hash = 0;
  } )"" <<
  Return( &ht[hash] );
\end{minted}
\caption{Cimple Hash Table lookup with linear probing.}
\label{lst:htlprobe}
\vspace{-8pt}
\end{listing}

\end{flushleft}

We compare performance of open-address hash table for
the special case of database hash-join.  
An ephemeral hash table optimized for hash-join~\cite{balkesen15tkde-modern} only needs to support bulk
insertion followed by a phase of lookups.  
The identity hash function can not be used in real workloads,
both for performance due to non-uniform skew, and for security due to
Denial-of-Service complexity attacks~\cite{crosby03sec-dos}.

\lstref{htlprobe} shows our classic linear probing hash table, similar to
the implementation suggested in Menon et al.~\cite{menon17vldb-fusion} --- linear probing at 50\% load factor, and Murmur3's
finalization as a hash, masked to the table size.  
They reported 1.2$\times$ gains from emitting via LLVM SIMD vectorization and using group
prefetching~\cite{chen04icde-join-prefetch}, on
their well-engineered hash table for state-of-the-art TPC-H performance.  

A requested static schedule for all three stages (delineated by \texttt{Yield} on lines 6 and 7 of \lstref{htlprobe}) generates three
independent static stages (shown in \appref{hash}).  Using the SoA layout enables compiler loop
vectorization to use AVX2 or AVX512
to calculate multiple hash functions simultaneously.

\paragraph{Variants}
Menon et al. ~\cite{menon17vldb-fusion}
analyze the inefficient baseline used in
AMAC~\cite{kocberber15vldb-amac}, i.e., identity hash, chaining at
400\% load factor, and using a linked list for handling duplicates.

For a chained hash table, which traverses a linked list, we can also produce a hybrid schedule.  The first two
steps use a static schedule (with SoA storage), while the third stage
can use a dynamic scheduler to handle a variable number of cache lines
traversed.

\section{Evaluation}
\label{sec:evaluation}

We report and analyze our performance gains
using Cimple used as a template library generator.
Our peak system throughput increases from
1.3$\times$ on HashTable to 2.5$\times$ on SkipList iteration; Cimple speedups of the time to complete a batch of queries
on a single thread range from 1.2$\times$ on HashTable to
6.4$\times$ on BinaryTree (\figref{speedup-search-intro}).

\subsection{Configuration}
\label{sec:hwcfg}
\paragraph{Hardware Configuration}

We used a dual socket Haswell system~\cite{ark-e5-v3-salike} with
24 cores at 2.9GHz clock frequency, or 3.3GHz for a single core.
Each socket has 4 memory
channels populated with dual rank, 16-bank DDR4-2133
memory~\cite{samsung-ddr4-2400}.  The DRAM memory level parallelism on
each socket therefore allows 128 open banks for random access.
The test applications were compiled with \texttt{gcc} 4.8 and executed on Linux 4.4 using huge pages.

\paragraph{Cimple Configuration}
\label{sec:swcfg}
We implemented the Cimple DSL in a combination of 2,500 lines of C++14, and
300 lines of C preprocessor macros. Cimple to C++ code was built with Apple clang 9.0.
The template library of runtime schedulers adds less than 500 lines of C++11 code.
Cimple suspend/resume of the minimum coroutine step (\textbf{SLi} -- 21 extra instructions) adds ~4ns.
We use 48 entry scheduler width --- optimal for all DRAM-resident benchmarks.

\subsection{Performance Gains}

\paragraph{Binary Search (BS)}
The multithreaded version has all threads searching from the same shared array of 1 billion 64-bit keys.
Branch-free execution is important for good performance as discussed in \secref{threekeys}.
When a branch is used on lines 10 and 11 of \lstref{bsearch}, we see only a 3$\times$ performance gain.  While CMOV in the baseline leads to a 0.7$\times$ slowdown,
Cimple+CMOV reaches 4.5$\times$ over the best baseline.

\paragraph{Binary Tree lookup (BT)}
Each thread works on a private tree to avoid synchronization, as used
in the context of partitioned single-threaded data stores, such as
VoltDB or Redis.  We use 1GB indexes scaled by the number of threads,
i.e., 48GB for the full system. 
We achieve 2.4$\times$ higher peak throughput, and
 6.4$\times$ speedup for a single thread of execution.
Our ability to boost a single thread performance much higher above
average, will support handling of skewed or bursty workloads, which
can otherwise cause significant degradation for partitioned
stores~\cite{taft2014store}.

\paragraph{SkipList lookup (SL)}

Concurrent SkipLists are much easier to scale and
implement~\cite{herlihy2006provably} compared to a binary tree,
therefore practical applications use multiple threads looking up items in a shared SkipList.

All items are found after a phase of insertions with no deletions, or other sources of memory fragmentation.
We achieve 2.7\mytimes{} single thread speedup, and 1.8\mytimes{} multithreaded throughput. 
Note that for SkipList lookup the ``down'' direction follows an array of pointers,
therefore the baseline benefits from speculative execution prefetching nodes.

\paragraph{SkipList Iterator (SLi)}
\label{sec:sliter}
We evaluated range queries on the same shared skip list index as above.
For 1,000 node limit iterations, similar to
long range queries in ~\cite{armstrong13sigmod-linkbench,sprenger2017book}
our total throughput gain is 2.5$\times$, and single thread speedup -- 4.1$\times$.

\paragraph{Hash Table lookup (HT)}
\label{sec:hteval}
Hash table join performance we evaluate on a table with 64-bit integer keys and values.
We use a 16~GB hash table shared among all threads, at an effective load factor of 48\%.  
We replicate similar single thread speedups~\cite{menon17vldb-fusion}
of 1.2$\times$ when either no results or all results are materialized.
Since there are few instructions needed to compare and store integer keys
and values, hardware is already very effective at keeping a high
number of outstanding requests.  However, both the hash table load
factor and the percentage of successful lookups impact branch
predictability, and thus ILP and MLP for the baseline.  For 50\%
materialized results, our speedup is 1.3$\times$.  When using 48 threads with 100\% hits, we get a 1.3$\times$ higher throughput of
650~M operations/s.

We also compared to other traditional but inefficient on modern cores variants, e.g., if
division by a prime number is used~\cite{chen04icde-join-prefetch} the
corresponding Cimple variant is 2$\times$ faster; and when there are
serializing instructions between lookups our speedup is 4$\times$.

\subsection{Performance Analysis}
\label{sec:perfcounters}

We analyze hardware performance counters to understand where our
transformations increase effective ILP and MLP.

\subsubsection{ILP Improvements}
\label{sec:evalstalls}

\tabref{mlpilp} shows our improvements in ILP and IPC by increasing the useful work per cycle,
and reducing the total number of cycles.
The ILP metric measures the average $\mu$instructions executed when not stalled (max 4).
Cimple may have either higher or lower instruction count: e.g., a pointer
dereference in \textbf{SLi} is a single instruction, while with a dynamic
scheduler that instruction is replaced by context switches with attendant register spills and restores.
For a static scheduler, vector instructions reduce additional instructions in \textbf{HT}.
The remaining stall cycles show that there is sufficient headroom for
more expensive computations per load.  

\subsubsection{MLP Improvements}
\label{sec:evalmlp}
Improving MLP lowers the stall penalty per miss, since up to 10
outstanding L1 cache misses per core can be overlapped.

In \tabref{mlpilp} we show that measured MLP improved by
1.3--6$\times$ with \CIMPLEnsp.  Measured as the average outstanding
L2 misses, this metric includes speculative and prefetch requests.
Therefore the baseline MLP may be inflated due to speculative
execution which does not always translate to performance.  Cimple
avoids most wasteful prefetching and speculation, therefore end-to-end
performance gains may be larger than MLP gains.

In BinarySearch, the baseline has high measured MLP due to speculation and prefetching, however, most of it is not contributing to effective MLP.
In BinaryTree the addresses of the children cannot be predicted, therefore the baseline has low MLP.
For SkipList lookup the down direction is an array of pointers therefore speculative execution may prefetch correctly needed values,
thus while the measured MLP is 2, the effective MLP is 1.5.
SkipList iteration is following pointers, and therefore has MLP of 1.
For HashTable at low load and 100\% hit rate, speculative execution is always correct, thus the baseline has high effective MLP.

There is also sufficient headroom in memory bandwidth and queue depth
for sequential input and output streams, e.g., for copying larger
payload values.

\begin{table}[t]
\footnotesize
  \centering
  \begin{tabular}{crccrccrccrc}
\textbf{Benchmark} & & \multicolumn{2}{c}{\textbf{MLP}} & &  \multicolumn{2}{c}{\textbf{ILP}} & &  \multicolumn{2}{c}{\textbf{IPC}} & &  \\
\cline{1-1}\cline{3-4}\cline{6-7}\cline{9-10}\noalign{\smallskip}
   &&  B & C && B & C  && B  & C && \\
BS && 7.5 & 8.5 && 1.6 & 2.3  && 0.13  & 1.10 &&  \\
BT && 1.2 & 4.3  && 1.6 & 2.3  &&  0.10 & 0.70   && \\
SL &&  2 & 5 && 1.8 & 2.4  && 0.07  & 0.60 && \\
SLi &&  1 & 5 && 1.3 & 2.0 && 0.01  &  0.22 && \\
HT &&  4.9 & 6.4 && 1.9 & 2.4  && 0.37  & 0.40 && \\
\noalign{\smallskip}
  \end{tabular}

  \caption{Memory Level Parallelism (MLP), Instruction Level Parallelism (ILP), and Instructions Per Cycle (IPC).  Baseline (B) vs Cimple (C). }
  \label{tab:mlpilp}
\vspace{-16pt}
\end{table}

\section{Related Work}
\label{sec:related}

\subsection{Hardware Multithreading}
Hardware context switching was explored in supercomputers of the
lineage of Denelcor HEP~\cite{smith86denelcor} and Tera
MTA~\cite{alverson90ics-tera}, e.g., Tera MTA supported 128
instruction streams that were sufficient to hide the latency of ~70
cycles of DRAM latency without using caches.
Yet locality is present in real workloads, and caches should be used
to capture different tiers of frequently used data.
Larrabee~\cite{seiler08siggraph-larrabee} threading and vectorization model allowed SIMD rebundling to maintain task efficiency.
Current GPUs offer large number of hardware threads, yet relying solely on
thread-level parallelism is insufficient~\cite{volkov10gtc-better},
and taking advantage of ILP and MLP is critical for GPU assembly-optimized
libraries~\cite{nervana17sgemm,lai13cgo-sgemm}.

Out-of-order CPUs can track the concurrent
execution of tens of co-running coroutines per core, but provide no
efficient notification of operation completion.
Informing loads~\cite{horowitz96isca-informing} were proposed as a
change to the memory abstraction to allow hardware to set a flag on a
cache miss and trap to a software cache-miss handler, similar to a
TLB-miss handler.  
Proposals for hardware support for overlapping instructions from different phases
of execution with compiler transformations have shown modest
performance gains\cite{tran17cgo-clairvoyance,sheikh12micro-cfd-dfd,ottoni05micro-dswp,raman08cgo-parallel-pipelining,vachharajani07pact-speculative}.

\subsection{Coroutines and Tasks}
Coroutines have been a low-level assembly technique since the 1950s,
originally used in limited-memory stackless environments ~\cite{newell56-IPL,newell60cacm-IPLV}. Lowering continuation overhead has been approached by specialized
allocation~\cite{hieb90pldi-corospace}, and partial~\cite{queinnec91popl-partial} or one-shot~\cite{bruggeman96pldi-oneshot} continuations.

The C++20 standard is also slated to support coroutines with the keywords
\verb|co_yield|, \verb|co_await|, and \verb|co_return|.  The original
proposals~\cite{cppstd-n3858-rf,cppstd-n4134-rf} motivate the goal to
make asynchronous I/O maintainable.  The runtime support for
completion tracking is acceptable at millisecond scale for handling
network and disk, but is too heavy-weight for tolerating nanosecond
memory delays targeted by \IMLP tasks.  The concise syntax and automatic state
capture are attractive, and 
the underlying mechanisms in LLVM and Clang can be used to add \Cimplesp as non-standard C++ extensions
to delight library writers.  Library users, however, can use the generated libraries
with standard compilers today.

Cilk~\cite{blumofe95cilk,frigo98pldi-cilk5} introduced an easy
programming model for fork-join task parallelism, divide-and-conquer
recursive task creation and work-stealing scheduler.  More recently
the CilkPlus~\cite{leiserson10jsc-cilkp} extensions to C/C++ were
added to \texttt{icc} and \texttt{gcc}.  C++20 proposals for
\verb|task_block|~\cite{cppstd-n4411-taskblock} incorporate task
parallelism like Cilk, albeit using a less succinct syntax.

Our rough guide to these overlapping programming
models would be to use C++20 tasks for compute bottlenecks, C++20 coroutines for
I/O, and \Cimplesp coroutines for memory bottlenecks.

\subsection{Software Optimizations}
Software prefetching by requesting data at a fixed distance ahead of
the current execution is complex even for simple loop nests and reference streams without
indirection ~\cite{mowry92asplos-prefetching}; and more recently surveyed in ~\cite{lee12taco-prefetch}.
Augmented data
structures help deeper pointer chasing~\cite{chilimbi99pldi-struct,kohout01pact}.

Optimized index data-structures for in-memory databases~\cite{rao00sigmod-csb,chen01sigmod-index,sewall11vldb-palm,
  krueger11vldb-merge,levandoski14sigmod-bwt-demo,adbirka16vldb-apollo, leis13icde-art,kemper13debull-hyper} try to reduce the depth of memory indirected references and use high fan-out and extra contiguous accesses
while performing one-at-a-time requests.
Techniques that uncover spatial or temporal locality by reordering
billions of memory
requests~\cite{kiriansky16pact-milk,beamer17ipdps-propagation} are
not applicable to index queries which often touch only a few rows.

Group prefetching and software-pipelined prefetching techniques were
introduced in~\cite{chen04icde-join-prefetch} where a group of
hash table lookups are processed as a vector, or with software pipelining; similarly used for an unordered key value store~\cite{li15isca-mica}.
AMAC~\cite{kocberber15vldb-amac} is
an extension to group prefetching to immediately refill completed tasks in order
to handle skewed inputs.  In a well-engineered baseline in the state-of-the-art database engine
Peloton~\cite{menon17vldb-fusion}, AMAC was found ineffective, and only group prefetching beneficial and maintainable.

Using C++20 coroutines for easier programmability of AMAC-style dynamic scheduling
was evaluated in concurrent work in SAP
HANA~\cite{psaropoulos17vldb-interleaving}.  While easier to use and
more maintainable than previous interleaving mechanisms
\cite{chen04icde-join-prefetch,kocberber15vldb-amac}, this resulted in
performance lower than \cite{chen04icde-join-prefetch}.
Dependence on the I/O-oriented coroutine implementation in Visual
C++ (MSVC) incurs high overheads; using a black-box
compiler is also not practical for JIT query engines used in modern databases for these critical inner loops.
For less critical code-paths implemented in C++, their promising
results are a step in the right direction.  We expect to be able to
offer a similar C++ front-end once coroutines are mature in Clang,
with additional Cimple AST annotations as C++ \verb|[[|attributes\verb|]]|.
Cimple's back-end seamlessly enables static and hybrid scheduling, with efficient
dynamic scheduling coroutines optimized for caches and out-of-order processors.

\section{Conclusion}
\label{sec:conclusion}

Cimple is fast, maintainable, and portable.  We offer an optimization
methodology for experts, and a tool usable by end-users today.

We introduced the \IMLP programming model for uncovering ILP and MLP
in pointer-rich and branch-heavy data structures and algorithms.  Our
\Cimplesp DSL and its AST transformations for C/C++ in \CIMPLEsp allow
quick exploration of high performance execution schedules.  Cimple
coroutine annotations mark hotspots with deep pointer dereferences or
long dependence chains.

Our compiler-independent DSL allows low-level programmers to generate
high-performance libraries that can be used by enterprise developers
using standard tool-chains.  Cimple performance results are faster than
all published manual optimizations, with up to 6.4$\times$ speedup.

\begin{acks}                            %

  This material is based upon work supported by
  DOE under Grant
  No.~\grantnum{GS100000001}{DE-SC0014204}, and
  the
  \grantsponsor{GS100000001}{Toyota Research Institute}
               {http://dx.doi.org/10.13039/100000001} under Grant
  No.~\grantnum{GS100000001}{LP-C000765-SR}.

Thanks to Carl Waldspurger for his helpful feedback on the presentation of earlier versions of this manuscript.

\end{acks}

\bibliography{milp}

\appendix
\section{Appendix}
\label{app:scheduler}

\subsection{Simplest Dynamic Scheduler}
\label{app:dynsimplest}

\begin{minted}[xleftmargin=10pt,gobble=4,fontsize=\footnotesize,linenos,escapeinside=~~]{c}

template<int Width, typename CoroutineState,
         typename Refill>
void SimplestScheduler(size_t tasks,
                       Refill refill
                       )
{
  CoroutineState cs[Width];
  // fill
  for(size_t i=0; i<Width; i++)
      refill(&cs[i], i);
  size_t nextTask = Width;
  while (nextTask < tasks) {
     for(size_t i=0; i<Width; i++) {
        if (cs[i].Step() && nextTask < tasks) {
            refill(&cs[i], nextTask);
            nextTask++;
        }
     }
  }
  // drain
  for(size_t i=0; i<Width; i++) {
     while (!cs[i].Done() && !cs[i].Step()) {
         // no further refill!
     }
  }
}
\end{minted}

\subsection{Push and Pull Callback Scheduler}
\label{app:pushpull}

\begin{minted}[xleftmargin=10pt,gobble=4,fontsize=\footnotesize,linenos,escapeinside=~~]{c}

template<int Width, typename CoroutineState,
         typename Push, typename Pull = NoPull>
void PushPullScheduler(Push push, Pull pull = Pull())
{
    CoroutineState cs[Width];
    bool draining = false;

    // fill
    for(size_t i=0; i<Width; i++)
        if (!push(&cs[i]))
            draining = true;
    
    int sup = 0;
    while (!draining) {
        for(size_t i=0; i<Width; i++) {
            if (cs[i].Step()) {
                if (!pull(&cs[i])) {
                    cs[i].Reset(); // generator
                } else if (!push(&cs[i]))
                    draining = true;
            }
        }
    }

    // drain
    for(size_t i=0; i<Width; i++) {
        while (!cs[i].Done() && !cs[i].Step()) {
            if ( !pull(&cs[i]) ) {
                cs[i]._state = 0;
            }
        }
    }
}
\end{minted}

\subsection{Coroutine and Static Schedule for Hash table lookup}
\label{app:hash}

\begin{minted}[xleftmargin=10pt,gobble=4,fontsize=\footnotesize,linenos,escapeinside=~~]{c}

// Do not edit by hand.
// Automatically generated with Cimple v0.1 \
from hashprobe_linear.cimple.cpp. 
// Compile with -std=c++11 

// Original HashTable_find
KeyType
HashTable_find(
	KeyType k)
{ 
   HashType hash;
     hash = Murmur3::fmix(k);
     hash &= mask;
     
  while (ht[hash].key != k &&
         ht[hash].key != 0) {
       hash++;
       if (hash == size) hash = 0;
  } 
     return ht[hash].key;
} 

// Coroutine state for HashTable_find
struct CoroutineState_HashTable_find { 
  CoroutineState_HashTable_find() : _state(0) {} 
  CoroutineState_HashTable_find(KeyType k) : 
	_state(0),
	k(k){ } 
  int _state = 0;
  KeyType _result;
  KeyType k;
  HashType hash;
  bool Done() const { return _state == _Finished; } 
  void Reset() { _state = 0; } 
  KeyType Result() const { return _result ; } 
  bool Step() { 
    switch(_state) {
    case 0:
        hash = Murmur3::fmix(k);
        hash &= mask;
        _state = 1;
        return false; 
    case 1:;
         // prefetch(&ht[hash]);
        _mm_prefetch((char*)(&ht[hash]),
                     _MM_HINT_T0);
        _state = 2;
        return false; 
    case 2:;
        
  while (ht[hash].key != k &&
         ht[hash].key != 0) {
       hash++;
       if (hash == size) hash = 0;
  } 
        _result = ht[hash].key;
        _state = _Finished;
        return true;
    } // switch 
    return false;
  } 
  constexpr static int InitialState = 0;
  constexpr static int _Finished = 3;
  enum class State { 
    Initial  = 0,
    Final  = 3
  }; 
};

// Coroutine SoA state for HashTable_find x 8
template <int _Width = 8>
struct CoroutineState_HashTable_find_8 { 
  int _state[_Width]  ;
  KeyType _result[_Width];
  KeyType _soa_k[_Width];
  HashType _soa_hash[_Width];
  bool SuperStep() { 
  for(int _i = 0; _i < _Width ; _i++) {
  KeyType& k = _soa_k[_i];
  HashType& hash = _soa_hash[_i];
        hash = Murmur3::fmix(k);
        hash &= mask;
        }
        for(int _i = 0; _i < _Width ; _i++) {
        KeyType& k = _soa_k[_i];
        HashType& hash = _soa_hash[_i];
         // prefetch(&ht[hash]);
        _mm_prefetch((char*)(&ht[hash]),
                     _MM_HINT_T0);
        }
        for(int _i = 0; _i < _Width ; _i++) {
        KeyType& k = _soa_k[_i];
        HashType& hash = _soa_hash[_i];
        
  while (ht[hash].key != k &&
         ht[hash].key != 0) {
       hash++;
       if (hash == size) hash = 0;
  } 
        _result[_i] = ht[hash].key;
  }
    return true; 
  } 
  void Init(  KeyType* k)  { 
     for(int _i=0; _i<_Width; _i++) { 
    _soa_k[_i] = k[_i];
     } 
  } 
  void Fini(KeyType*out)  {  
     for(int _i=0; _i<_Width; _i++) { 
        out[_i] = _result[_i]; 
     } 
  } 
 }; 
// End of batched HashTable_find
\end{minted}

\end{document}